\documentclass{article}

\usepackage{spconf,amsmath,epsfig}
\usepackage{times,amsmath,epsfig}
\usepackage{latexsym,amssymb}
\usepackage{cite}

\newcommand{\thetav}{\hbox{\boldmath$\theta$}}

\renewcommand{\arg}{{\hbox{arg}}}

\newcommand{\beq}{\begin{equation}}
\newcommand{\enq}{\end{equation}}
\newcommand{\beqa}{\begin{eqnarray}}
\newcommand{\enqa}{\end{eqnarray}}

\newcommand{\beql}[1]{\begin{equation}\label{#1}}

\newcommand{\no}{\nonumber}

\newtheorem{lem}{Lemma}

\newtheorem{rl}{Rule}
\begin{document}

\title{OPTIMAL MEDIUM ACCESS CONTROL IN COGNITIVE RADIOS: A SEQUENTIAL DESIGN APPROACH}
\name{Lifeng Lai$^1$\thanks{This research was supported by the
National Science Foundation under Grants ANI-03-38807 and
CNS-06-25637.}, Hesham El Gamal$^2$, Hai Jiang$^3$ and H. Vincent
Poor$^1$}
\address{$^1$Dept. of Elec. Eng., Princeton Univ.,  \{llai, poor\}@princeton.edu\\
$^2$Dept. of Elec. and Comp. Eng., Ohio State Univ.,  helgamal@ece.osu.edu\\
$^3$Dept. of Elec. and Comp. Eng., Univ. of Alberta,
hai.jiang@ece.ualberta.ca}

\date{}

\maketitle 



\begin{abstract}
The design of medium access control protocols for a cognitive user
wishing to opportunistically exploit frequency bands within parts of the radio spectrum having multiple bands is considered. In the scenario under consideration, the availability probability of each channel is unknown \emph{a priori} to the cognitive user. Hence efficient medium access strategies must strike a balance between exploring the availability of channels and exploiting the
opportunities identified thus far. Using a sequential design
approach, an optimal medium access strategy is derived. To avoid
the prohibitive computational complexity of this optimal strategy,
a low complexity asymptotically optimal strategy is also
developed. The proposed strategy does not require any
prior statistical knowledge about the traffic pattern on the
different channels.
\end{abstract}
\vspace{-2mm}
\section{Introduction} \label{sec:intro}
As a promising technique to increase spectral efficiency of
overcrowded parts of the radio spectrum, the opportunistic spectrum
access problem has been the focus of significant research
activities~\cite{Mitola:IPC:99}. The underlying idea is to allow
unlicensed users (i.e., cognitive users) to access the available
spectrum when the licensed users (i.e., primary users) are not
active. The presence of high priority primary users and the
requirement that the cognitive users should not interfere with
them introduce new challenges for protocol design. The overarching
goal of the current work is to develop a unified framework for the design of efficient, and low complexity, cognitive medium access
protocols.

The spectral opportunities available to cognitive users are by their nature time-varying. To avoid interfering with the primary
network, cognitive users must first probe to determine whether
there are primary activities before transmission.
Under the assumption that each cognitive user cannot access all of
the available channels simultaneously, the main task of the medium
access protocol is to distributively choose which channels each
cognitive user should attempt to use in different time slots, in
order to fully (or maximally) utilize the spectral opportunities.
This decision process can be enhanced by taking into account any
available statistical information about the primary traffic. For
example, with a single cognitive user capable of accessing
(sensing) only one channel at a time, the problem becomes trivial
if the probability that each channel is free is known {\em a
priori}. In this case, the optimal rule is for the cognitive user
to access the channel with the highest probability of being free
in all time slots. However, such time-varying traffic information
is typically not available to the cognitive users {\em a priori}.
The need to learn this information on-line creates a fundamental
tradeoff between exploitation and exploration. Exploitation refers
to the short-term gain resulting from accessing the channel with
the estimated highest probability of being free (based on the
results of previous sensing decisions) whereas exploration is the
process by which a cognitive user learns the statistical
behavior of the primary traffic (by choosing possibly different
channels to probe across time slots). In the presence of multiple
cognitive users, the medium access algorithm must also account for
the competition between different users over the same channel.

In this paper, we develop a unified framework for the design and
analysis of cognitive medium access protocols in the presence of a
single cognitive user who can access a single channel in each time slot. As argued in the sequel, this framework allows for the construction of strategies that strike an optimal balance between exploration and exploitation. We derive an optimal sensing rule that maximizes the expected throughput obtained by the cognitive user. Compared with a genie-aided scheme, in which the cognitive user knows {\em a priori} the primary network traffic information, there is a throughput loss suffered by any medium access strategy. We obtain a lower bound on this loss and further construct a linear
complexity single index protocol that achieves this lower bound
asymptotically (when the primary traffic behavior changes slowly).
Similar approaches have been considered in~\cite{Motamedi:ETTRT:07}
and~\cite{Zhao:JSAC:07}, but with different emphases.

We have also extended our study to networks with multiple
cognitive users and networks with more capable cognitive users, and have developed optimal strategies for these scenarios. However, due to space limitations, we do not discuss these results here. We also omit the proofs of results presented in this paper. Interested readers can refer to~\cite{Lai:TON:07} for details.

The rest of this paper is organized as follows. Our network model
is detailed in Section~\ref{sec:model}. Section~\ref{sec:full}
develops and analyzes an optimal strategy for the single cognitive
user scenario. Finally, Section~\ref{sec:con} summarizes our
conclusions.
\vspace{-2mm}
\section{Network Model}\label{sec:model}

\begin{figure}[thb]
\centering
\includegraphics[width=0.45 \textwidth]{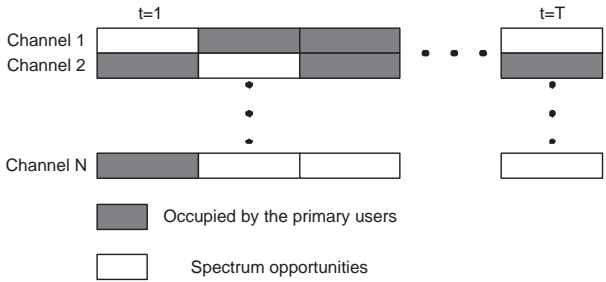}
\caption{Channel model.} \label{fig:model}
\end{figure}

Figure~\ref{fig:model} shows the channel model of interest. We
consider a primary network consisting of $N$ non-overlapping channels, $\mathcal{N}=\{1,\cdots,N\}$, each with bandwidth $B$. The users
in the primary network are operated in a synchronous time-slotted
fashion. We assume that at each time slot, channel $i$ is free
with probability $\theta_i$. Let $Z_{i}(j)$ be a random variable
that equals $1$ if channel $i$ is free at time slot $j$ and equals
$0$ otherwise. Hence, given $\theta_i$, $Z_{i}(j)$ is a Bernoulli
random variable with distribution
$h_{\theta_i}(z_{i}(j))=\theta_i\delta(1)+(1-\theta_i)\delta(0),$
where $\delta(\cdot)$ is a delta function. Furthermore, for a
given $\thetav=(\theta_1,\cdots,\theta_N)$, the $Z_{i}(j)$ are
independent for each $i$ and $j$. We consider a block varying
model in which the value of $\thetav$ is fixed for a block of $T$
time slots and then randomly changes at the beginning of the next block
according to a joint probability density function (pdf)
$f(\thetav)$. \vspace{-1mm}
\vspace{-1mm}

In our model, the cognitive users attempt to exploit the
availability of free channels in the primary network by sensing
the activity at the beginning of each time slot. Our work seeks to
characterize efficient strategies for choosing which channels to
sense (access). The challenge here stems from the fact that the
cognitive users are assumed to be unaware of $\thetav$ {\em a
priori}. We consider two cases in which a cognitive user either
has or does not have prior information about the pdf of $\thetav$,
i.e., $f(\thetav)$. In the scenario presented in this paper, at
time slot $j$, a single cognitive user selects one channel
$S(j)\in\mathcal{N}$ to access. If the sensing result shows that
channel $S(j)$ is free, \emph{i.e.}, $Z_{S(j)}(j)=1$, the
cognitive user can send $B$ bits over this channel; otherwise, the
cognitive user will wait until the next time slot and pick a
possibly different channel to access. Therefore, the total number
of bits that the cognitive user is able to send over one block (of
$T$ time slots) is
$$W=\sum\limits_{j=1}^{T}BZ_{S(j)}(j).$$

It is clear that $W$ is a random variable that depends on the
traffic in the primary network and, more importantly for us, the
medium access protocols employed by the cognitive user. Therefore,
the overarching goal of this paper is to construct low
complexity medium access protocols that maximize
$\mathbb{E}\{W\}$.

Intuitively, the cognitive user would like to select the channel
with the highest probability of being free in order to obtain more
transmission opportunities. If $\thetav$ is known then this
problem is trivial: the cognitive user should choose the channel
$i^*=\arg\max\limits_{i\in\mathcal{N}}\theta_i$ to sense. The
uncertainty in $\thetav$ imposes a fundamental tradeoff between
exploration, in order to learn $\thetav$, and exploitation, by
accessing the channel with the highest estimated free probability
based on current available information, as detailed in the
following section. \vspace{-2mm}
\section{Optimal Medium Access Protocols}\label{sec:full}
We start by developing the optimal solution under the idealized
assumption that $f(\thetav)$ is known {\em a priori} by the
cognitive user. As we will see, this optimal medium access
algorithm suffers from a prohibitive computational complexity that
grows exponentially with the block length $T$. This motivates the
design of low complexity asymptotically optimal approaches, which
we also consider.

Our cognitive medium access problem
belongs to the class of bandit problems. In this setting, the
decision maker must sequentially choose one process to observe
from $N\geq 2$ stochastic processes. These processes usually have
parameters that are unknown to the decision maker and, associated
with each observation is a utility function. The objective of the
decision maker is to maximize the sum or discounted sum of the
utilities via a strategy that specifies which process to observe
for every possible history of selections and observations. A
comprehensive treatment covering different variants of bandit
problems can be found in~\cite{Berry:Book:85}.


We are now ready to rigorously formulate our problem. The
cognitive user employs a medium access strategy $\Gamma$, which
will select channel $S(j)\in\mathcal{N}$ to sense at time slot $j$
for any possible causal information pattern obtained through the
previous $j-1$ observations:
$\Psi(j)=\{s(1),z_{s(1)}(1),\cdots,s(j-1),z_{s(j-1)}(j-1)\}, j\geq
2,$ \emph{i.e.} $s(j)=\Gamma(f,\Psi(j))$. Notice that
$z_{s(j)}(j)$ is the sensing outcome of the $j^{th}$ time slot, in
which $s(j)$ is the channel being accessed. If $j=1$, there is no
accumulated information, and thus $\Psi(1)=\phi$ and $s(1)=\Gamma(f)$.
The utility that the cognitive user obtains by making decision
$S(j)$ at time slot $j$ is the number of bits it can transmit at
time slot $j$, which is $BZ_{S(j)}(j)$. We denote the expected
value of the payoff obtained by a cognitive user who uses strategy
$\Gamma$ as \vspace{-1mm}
\begin{eqnarray}\label{eq:prob}
W_{\Gamma}=\mathbb{E}_{f}\left\{\sum
\limits_{j=1}^{T}BZ_{S(j)}(j)\right\}.
\end{eqnarray}
\vspace{-1mm}

We further denote $V^*(f,T)=\sup\limits_{\Gamma}W_{\Gamma}$, which is the
largest throughput that the cognitive user could obtain when the
spectral opportunities are governed by $f(\thetav)$ and the exact
value of each realization of $\thetav$ is not known \emph{a priori} by
the user.

Each medium access decision made by the cognitive user has two
effects. The first one is the short-term gain, i.e., an immediate
transmission opportunity if the chosen channel is found free. The
second one is the long-term gain, i.e., the updated statistical
information about $f(\thetav)$. This information will help the
cognitive user in making better decisions in future stages.
There is an interesting tradeoff between the short and long-term
gains. If we only want to maximize the short-term gain, we can
choose the channel with the highest availability probability to sense, based on
the current information. This myopic strategy maximally exploits
the existing information. On the other hand, by choosing other
channels to sense, we gain statistical information about
$f(\thetav)$ which can effectively guide future decisions. This
process is typically referred to as exploration, as noted previously.

More specifically, let $f^{j}(\thetav)$ be the updated pdf after
making $j-1$ observations. We begin with
$f^{1}(\thetav)=f(\thetav)$. After observing $z_{s(j)}(j)$, we
update the pdf using the following Bayesian formula.
\vspace{-1mm}
\begin{enumerate}
\item If $z_{s(j)}(j)=1$,
$f^{j+1}(\thetav)=\frac{\theta_{s(j)}f^{j}(\thetav)}{\int\theta_{s(j)}f^j(\thetav)d\thetav},$
\item If $z_{s(j)}(j)=0$,
$f^{j+1}(\thetav)=\frac{\left(1-\theta_{s(j)}\right)f^{j}(\thetav)}{\int\left(1-\theta_{s(j)}\right)f^j(\thetav)d\thetav}.$
\end{enumerate}
The following result characterizes the optimal strategy that
maximizes the average throughput the cognitive user obtains from
the network.
\vspace{-2mm}
\begin{lem}\label{lem:optimal}
For any prior pdf $f$, there exists an optimal strategy $\Gamma^*$
to the channel selection problem~\eqref{eq:prob}, and $V^*(f,T)$
is achievable. Moreover, $V^*$ satisfies the following condition:
\begin{eqnarray}\label{eq:solution}
V^*(f,T)=\max\limits_{s(1)\in\mathcal{N}}\mathbb{E}_{f}\left\{BZ_{s(1)}+V^*\left(f_{Z_{s(1)}},T-1\right)\right\},
\end{eqnarray}
where $f_{Z_{s(1)}}$ is the conditional distribution updated using
the Bayesian rule described above, as if the cognitive user
chooses $s(1)$ and observes $Z_{s(1)}$. Also,
$V^*\left(f_{Z_{s(1)}},T-1\right)$ is the value of a bandit
problem with prior information $f_{Z_{s(1)}}$ and $T-1$ sequential
observations.
\end{lem}
\vspace{-2mm}

In principle, Lemma~\ref{lem:optimal} provides the solution to
problem~\eqref{eq:prob}. Effectively, it decouples the calculation
at each stage, and hence, allows the use of dynamic programming to
solve the problem. The idea is to solve the channel selection
problem with a smaller dimension first and then use backward
deduction to obtain the optimal solution for a problem with a
larger dimension. Starting with $T=1$, the second term inside the
expectation in~\eqref{eq:solution} is 0, since $T-1=0$. Hence, the
optimal solution is to choose the channel $i$ having the largest
$\mathbb{E}_f\{BZ_{i}\}$, which can be calculated as
$\mathbb{E}_f\{BZ_{i}\}=B\int \theta_i f(\thetav)d\thetav.$
And $V^*(f,1)=\max\limits_{i\in\mathcal{N}}\mathbb{E}_f\{BZ_i\}$.
With the solution for $T=1$ at hand, we can now solve the $T=2$
case using~\eqref{eq:solution}. At first, for every possible
choice of $s(1)$ and possible observation $z_{s(1)}$, we calculate
the updated distribution $f_{z_{s(1)}}$ using the Bayesian
formula. Next, we calculate $V^*(f_{z_{s(1)}},1)$ (which is
equivalent to the $T=1$ problem described above). Finally,
applying~\eqref{eq:solution}, we have the following equation for
the channel selection problems with $T=2$: \vspace{-1mm}
\begin{eqnarray}
V^*(f,2)&=&\max\limits_{i\in \mathcal{N}}\int
\left[B\theta_i+\theta_i
V^{*}(f_{z_i=1},1)\right.\no\\&&\hspace{12mm}\left.+(1-\theta_i)V^*(f_{z_i=0},1)\right]f(\thetav)d\thetav.\no
\end{eqnarray}
Hence, in the first step, the cognitive user should choose
$i^*(1)=\arg\max\limits_{i\in\mathcal{N}}V^*(f,2)$ to sense. After
observing $z_{i^*(1)}$, the cognitive user has
$\Psi(1)=\{z_{i^*(1)}\}$, and it should choose
$i^*(2)=\arg\max\limits_{i\in\mathcal{N}}V^*(f_{z_{i^*(1)}},1)$.
Similarly, after solving the $T=2$ problem, one can proceed to
solve the $T=3$ case. Using this procedure recursively, we can
solve the problem with $T-1$ observations. Finally, our original
problem with $T$ observations is solved as follows. \vspace{-1mm}
\begin{eqnarray}
V^*(f,T)\hspace{-2mm}&=\hspace{-2mm}&\max\limits_{i\in \mathcal{N}}\int
\left[B\theta_i+\theta_i V^{*}(f_{z_i=1},T-1)\right.\no\\
&&\hspace{12mm}\left.+(1-\theta_i)V^*(f_{z_i=0},T-1)\right]f(\thetav)d\thetav.\no
\end{eqnarray}
\vspace{-2mm}

The optimal solution developed above suffers from a prohibitive
computational complexity. In particular, the dimensionality of our
search dimension grows exponentially with the block length $T$.
Moreover, one can envision many practical scenarios in which it
would be difficult for the cognitive user to obtain the prior
information $f(\thetav)$. This motivates our pursuit of low
complexity non-parametric protocols which maintain certain
optimality properties and do not depend on $f(\thetav)$
explicitly. Hence, in the following, we aim to develop strategies
that depend only on the information obtained through observations
$\Psi$.

For a given strategy $\Gamma$, the expected number of bits the
cognitive user is able to transmit through a block with given
parameters $\thetav$ is
\vspace{-1mm}
\begin{eqnarray}
\mathbb{E}\left\{\sum \limits_{j=1}^{T}BZ_{S(j)}(j)\right\}
=\sum\limits_{j=1}^{T}B\sum\limits_{i=1}^N\theta_i
\text{Pr}\left\{\Gamma(\Psi(j))=i\right\}.\no
\end{eqnarray}
\vspace{-1mm}

Recall that $\Gamma(\Psi(j))=i$ means that, following strategy
$\Gamma$, the cognitive user should choose channel $i$ in time
slot $j$, based on the available information $\Psi(j)$. Here
$\text{Pr}\left\{\Gamma(\Psi(j))=i\right\}$ is the probability
that the cognitive user will choose channel $i$ at time slot $j$,
following the strategy $\Gamma$.

Compared with the idealistic case where the exact value of
$\thetav$ is known, in which the optimal strategy for the
cognitive user is to always choose the channel with the largest
availability probability, the loss incurred by $\Gamma$ is given by
\vspace{-1mm}
\begin{eqnarray}
L(\thetav;\Gamma)=\sum\limits_{j=1}^TB\theta_{i^*}-\sum\limits_{j=1}^{T}B\sum\limits_{i=1}^N\theta_i
\text{Pr}\left\{\Gamma(\Psi(j))=i\right\},\no
\end{eqnarray}
where $\theta_{i^*}=\max\{\theta_1,\cdots,\theta_N\}$. We say that
a strategy $\Gamma$ is consistent if, for any $\thetav\in[0,1]^N$,
there exists $\beta<1$ such that $L(\thetav;\Gamma)$ scales as
$O(T^{\beta})$. In the sequel, we use the following notations 1)
$g_1(N)=\omega(g_2(N))$ means that $\forall c>0,\exists N_{0},$ such that $\forall
N>N_{0}, g_2(N)< c g_1(N) $; 2) $g_1(n)=O(g_2(N))$ means that $\exists
c_{1},c_{2}>0$ and $N_{0}$, such that $\forall N>N_{0},$ $ c_{1}g_2(N)\leq
g_1(N)\leq c_{2}g_2(N)$. For example, consider a loyal scheme in
which the cognitive user selects channel $i$ at the beginning of a
block and sticks to it. If $\theta_i$ is the largest one among
$\thetav$, $L(\thetav;\Gamma)=0$. On the other hand, if $\theta_i$
is not the largest one, $L(\thetav;\Gamma)\sim O(T)$. Hence, this
loyal scheme is not consistent. The following lemma characterizes
the fundamental limits of any consistent scheme.

\begin{lem}\label{lem:lowerbound} For any $\thetav$ and any consistent strategy $\Gamma$, we have
\begin{eqnarray}
\lim\inf\limits_{T\rightarrow\infty} \frac{L(\thetav;\Gamma)}{\ln
T}\geq B\sum \limits_{i\in\mathcal{N}\backslash
\{i^*\}}\frac{\theta_{i^*}-\theta_i}{D(\theta_i||\theta_i^*)},
\end{eqnarray}
where $D(\theta_i||\theta_l)$ denotes the Kullback-Leibler divergence
between the two Bernoulli random variables with parameters
$\theta_i$ and $\theta_l$ respectively:
$D(\theta_i||\theta_l)=\theta_i\ln\left(\frac{\theta_i}{\theta_l}\right)+(1-\theta_i)\ln\left(\frac{1-\theta_i}{1-\theta_l}\right).$
\end{lem}
\vspace{-1mm} Lemma~\ref{lem:lowerbound} shows that the loss of
any consistent strategy scales at least as $\omega(\ln T)$. An
intuitive explanation of this loss is that we need to spend at
least $O(\ln T)$ time slots on sampling each of the channels with
smaller $\theta_i$, in order to get a reasonably accurate estimate
of $\thetav$, and hence use it to determine the channel having
the largest $\theta_i$ to sense. We say that a strategy $\Gamma$
is order optimal if $L(\thetav;\Gamma)\sim O(\ln T)$.

Before proceeding to the proposed low complexity order-optimal
strategy, we first analyze the loss order of some heuristic
strategies which may appear to be reasonable.

The first simple rule is the random strategy $\Gamma_r$ where, at
each time slot, the cognitive user randomly chooses a channel from
the available $N$ channels. The fraction of time the
cognitive user spends on each channel is therefore $1/N$, leading
to the loss $L(\thetav;\Gamma_r)=\frac{B\sum\limits_{i=1}^N
(\theta_{i^*}-\theta_i)}{N}T\sim O(T).$

The second one is the myopic rule $\Gamma_g$ in which the
cognitive user keeps updating $f^j(\thetav)$, and chooses the
channel with the largest value of $\hat{\theta}_i=\int\theta_i
f^j(\thetav)d\thetav$ at each stage. Since there are no convergence
guarantees for the myopic rule, that is $\hat{\thetav}$ may never
converge to $\thetav$ due to the lack of sufficiently many samples
for each channel~\cite{Kumar:SJCO:85}, the loss of this myopic
strategy is $O(T)$.

The third protocol we consider is {\em staying with the winner and
switching from the loser rule} $\Gamma_{SW}$ where the cognitive
user randomly chooses a channel in the first time slot. In the
succeeding  time-slots 1) if the accessed channel was found to be
free, it will choose the same channel to sense; 2) otherwise, it
will choose one of the remaining channels based on a certain
switching rule. \vspace{-1mm}
\begin{lem}\label{lem:sw}
No matter what the switching rule is, $L(\thetav;\Gamma_{SW})\sim
O(T)$.
\end{lem}
\vspace{-1mm} There are several strategies that have loss of order
$O(\ln T)$. We adopt the following linear complexity strategy
from~\cite{Auer:ML:02}.

\begin{rl}(Order optimal  single index strategy)

The cognitive user maintains two vectors $\mathbf{X}$ and
$\mathbf{Y}$, where each $X_i$ records the number of time slots
in which the cognitive user has sensed channel $i$ to be free,
and each $Y_i$ records the number of time slots in which the
cognitive user has chosen channel $i$ to sense. The strategy works
as follows.

\begin{enumerate}

\item Initialization: at the beginning of each block, each
channel is sensed once.

\item After the initialization period, the cognitive user obtains
an estimate $\hat{\thetav}$ at the beginning of time slot $j$,
given by
$\hat{\theta}_i(j)=X_i(j)/Y_i(j),$
and assigns an index $\Lambda_i(j)=\hat{\theta}_i(j)+\sqrt{2\ln
j/Y_i(j)}$ to the $i^{th}$ channel. The cognitive user chooses the
channel with the largest value of $\Lambda_i(j)$ to sense at time
slot $j$. After each sensing, the cognitive user updates
$\mathbf{X}$ and $\mathbf{Y}$.
\end{enumerate}
\end{rl}
\vspace{-2mm}

The intuition behind this strategy is that as long as $Y_i$ grows
as fast as $O(\ln T)$, $\Lambda_i$ converges to the true value of
$\theta_i$ in probability, and the cognitive user will choose the
channel with the largest $\theta_i$ eventually. The loss of $O(\ln
T)$ comes from the time spent in sampling the inferior channels in
order to learn the value of $\thetav$. This price, however, is
inevitable as established in the lower bound of
Lemma~\ref{lem:lowerbound}.

Finally, we observe that the difference between the myopic rule
and the order optimal single index rule is the additional term
$\sqrt{2\ln j/Y_i(j)}$ added to the current estimate
$\hat{\theta}_i$. Roughly speaking, this additional term
guarantees enough sampling time for each channel, since if we
sample channel $i$ too sparsely, $Y_i(j)$ will be small, which
will increase the probability that $\Lambda_i$ is the largest
index. When $Y_i(j)$ scales as $\ln T$, $\hat{\theta}_i$ will be
the dominant term in the index $\Lambda_i$, and hence the channel
with the largest $\theta_i$ will be chosen much more frequently.

\vspace{-3mm}

\section{Conclusions}\label{sec:con}
This work has developed a unified framework for the design and
analysis of cognitive medium access based on the classical bandit
problem. Our formulation highlights the tradeoff between
exploration and exploitation in cognitive channel selection. A
linear complexity cognitive medium access algorithm, which is
asymptotically optimal as the number of time slots increases, has also been
proposed.

\end{document}